\documentclass[secnumarabic,amssymb, nobibnotes,]{article}

\usepackage{amsmath}
 \usepackage{amsmath}
    \usepackage{amssymb}
    \usepackage{amsthm}
    \usepackage{amsfonts}
    \usepackage{braket}

\usepackage{color}
\usepackage{pstricks}
\usepackage{graphicx}
\usepackage{slashed}

\begin{document}

\title{On the infrared divergence and global colour in ${\mathcal N}=4$ Yang-Mills theory}

\author{ Su Yu Ding, Joanna Karczmarek,  Gordon W. Semenoff 
\\~\\
 Department of Physics and
Astronomy, University of British Columbia,\\ 6224 Agricultural Road,
Vancouver, BC Canada V6T 1Z1 }

 \maketitle

\begin{abstract}
The ${\mathcal N}=4$ superconformal Yang-Mills theory on flat four-dimensional Minkowski space is a de-confined gauge theory in the sense that the string tension for fundamental representation coloured quarks vanishes. In fact, static fundamental representation quarks which lie in certain half-BPS super-multiplets do not interact at all.  An interesting question asks whether such quarks would carry a well-defined global colour charge which, when the gauge is fixed, should have the status of an internal symmetry.  We shall present a simple paradigmatic model which suggests that the answer to this question lies in the way in which infrared divergences are dealt with.  

   \end{abstract}

\section{Introduction}

There is a fundamental issue as to whether the $S$ matrix exists and whether it can be a useful tool for the study of  four dimensional quantum field theories where the interactions are mediated by massless fields.    Such interactions do not decouple sufficiently rapidly at large distances and times to justify the assumption that particles become free so that the $S$-matrix can be used to describe transitions between free particle states.  If one ignores the problem and proceeds to compute amplitudes using the standard diagrammatic perturbation theory, the difficulty is manifest as  infrared and co-linear divergences.    

In Abelian gauge theories such as quantum electrodynamics, techniques for dealing with  infrared divergences are well known.  There are two principal approaches, either the computation of transition probabilities inclusive of soft photon production  \cite{Bloch:1937pw}-\cite{Jauch:1976ava} or the use of the $S$ matrix to compute transition amplitudes between 
dressed states  \cite{Chung:1965zza}-\cite{Kulish:1970ut}. 
They give identical results for transition probabilities.  
 However, it has recently been pointed out \cite{Carney:2017jut}-\cite{Semenoff:2019dqe} that the two approaches have subtle differences with how quantum information is distributed in a scattering process \cite{Park:2014hya}-\cite{Araujo:2019mni}.  
  
 The infrared problem in non-Abelian gauge theory is significantly more difficult than that of its Abelian cousin.   
 The Kinoshita-Lee-Nauenberg theorem \cite{Kinoshita:1962ur}-\cite{Lee:1964is} guarantees that the probabilities of completely inclusive processes must
 be infrared finite.  By ``completely inclusive'' we mean that one must sum over both all possible final states and all possible initial states containing soft particles.   The sum over initial states is particularly subtle.   It has ambiguous normalization, which some effort has gone into treating systematically \cite{Akhoury:1997pb}-\cite{Khalil:2017yiy}.  Having to sum over initial states is tantamount to declaring that the $S$-matrix can only be applied to incoming mixed states where the density matrix has equal probability of finding any of the possible incoming states, including those with arbitrary numbers of soft particles.  Fortunately, there is recent work that suggests that this limitation is not necessary, that it is sufficient to sum over final states alone (or initial states alone)  if forward scattering is properly taken into account \cite{Frye:2018xjj}. This would indeed be a very useful development and it is entirely consistent with the results that we will find in the following. 
 There has also been a certain amount of work on the alternative of attempting to find the analog of the dressed states for a non-Abelian gauge theory \cite{Butler:1978rd}-\cite{Gonzo:2019fai}.   We will also make use of these in the following, but only in the leading perturbative order where they resemble the dressed states of Abelian gauge theory.
  
 In this paper we will observe another major difference between the inclusive and dressed approaches to the infrared problem
 in the context of a simple perturbative computation of a particler amplitude in ${\mathcal N}=4$ supersymmetric Yang-Mills theory.
Our result will illustrate that the way in which infrared singularities are dealt with impinges on the question as to whether it makes sense, in a gauge fixed field theory with de-confined quarks, to assign a specific colour to a quark.  
 ${\mathcal N}=4$  Yang-Mills theory is
 a super-conformal gauge theory where it is known that the string tension for a certain type of heavy quark vanishes, meaning that quarks  are not confined.  In fact, the quarks that we are interested in are members of a $\frac{1}{2}$-BPS super-multiplet and some of  their properties are protected by supersymmetry.  For example, static quarks do not interact at all.  
 
The quarks that we will study transform in the fundamental representation of a U(N) gauge group.  A quark wave-function is therefore an $N$ component object where, in some canonical basis, each component corresponds to a colour.  We would say that the quark has definite colour when only one component is non-zero.  Then, we can ask the question as to whether it makes any physical sense to think of the quark as being in such a state of definite colour.  Of course, we can only ask this question once we have fixed a gauge by choosing a representative of each gauge orbit.  Once that is done, there is a global SU(N) symmetry and we expect that the states of the quantum field theory can be organized so that they carry representations of that symmetry group.  A state with a single quark would transform in the $N$-dimensional fundamental representation. 

We will consider the amplitude that a constantly accelerating quark with a given colour evolves classically, that is, the probability amplitude that it follows its classical trajectory and no other particles are produced.  We will  take the heavy quark limit so that we can analyze the problem in a semi-classical expansion.   We will also take the limit where the time interval during which the constant acceleration occurs is large.  For a free quark, the semiclassical limit of this amplitude is simply a phase ${\mathcal A}_{ab}\sim e^{-i\delta\tau_P}\delta_{ab}$, with $\tau_P$ the proper time during which the quark accelerates and $\delta$ the particle mass, perhaps corrected by the particle's interaction with the accelerator.  This is simply the statement that, in the classical limit, the particle is sure to follow its classical trajectory.  The indices $a$ and $b$ are quark colours and the delta function indicates that the colour of the quark is unchanged during its propagation.  

We will then compute the corrections to the amplitude from the coupling of the quark to the massless fields of ${\mathcal N}=4$ Yang-Mills theory, to the leading order of perturbation theory. We will find that the leading correction is quadratic, rather than linear in the proper time of the quark, a fact that we attribute to an infrared divergence that would occur if the proper time were taken to infinity. 

We will study two ways in which this infrared problem can potentially be solved.  The first is the inclusive approach. It considers a process which is inclusive of
the production of soft vector and scalar particles.  The second uses dressed states.  It uses the conventional S-matrix to study the transitions between dressed states consisting of on-shell heavy quarks and   coherent states of the on-shell massless vector and scalar particles of ${\mathcal N}=4$ Yang-Mills theory.

The result of our first attempt might be quite obvious from the beginning.   When a coloured quark accelerates, it emits  bremsstrahlung which includes soft vector and scalar fields.   The soft particles that are emitted are arbitrarily soft, to the point where they are not noticeable in the kinematics of the process.  However, they carry colour charges and each time the quark emits such a particle  its colour state flips.   The soft vector and scalar particles transform in the adjoint representation and their emission is equally capable of flipping any fundamental representation quark colour to any other colour.  This continuous flipping of colour, every time the quark accelerates, and the loss of this information to the soft particles which escape detection,  randomizes the colour of the original quark.   In this picture, any quark which has ever been accelerated cannot be in a definite state of colour, but it must be in a mixed state where all of the colours are weighted equally. 

Our second result tells us that we can redefine what we mean by a quark by adding a cloud of soft on-shell vector and scalar particles to make a dressed state.  We shall do this only to the leading order of perturbation theory.  In the dressed state, the quark together with the soft particles combine to transform in the fundamental representation of colour SU(N).  The dressing is tuned so that the infrared divergences cancel.  Then, the infrared  decouples in that the evolution of the system through transitions between dressed states involves no additional production of soft particles.    As a result, as long as coloured hard particles are not produced by an interaction, the quark retains its colour.  Its colour no longer fluctuates and the colour state seems to be a meaningful attribute of the quark.  However,  inside the dressed state, colour is distributed in a nonlocal way between the  quark and the soft  particles,  which must combine so that they are altogether in the fundamental representation.  They do this in such a way that
the colour state of the quark and the colour state of the soft particles are maximally entangled, even when they are very far separated. This fact suggests that, even in the dressed state formalism, colour is not a local observable.
 
In section 2 we will review the setup of the heavy quark scenario in ${\mathcal N}=4$ supersymmetric Yang-Mills theory. 
 In section 3 we will analyze a constantly accelerating heavy quark which emits a soft gluon.  It is there that we show that the inclusion of the possibility of soft particle production can repair the infrared divergence.  This is so only for certain incoming or outgoing states, those which are mixed states in which colour is completely random. 
 In those states, the quantum correction to the amplitude that the quark follows it classical trajectory becomes a linear, rather than a quadratic in the proper time of the quark.  We extract a ``damping rate'' from this linear proper time dependence.  (In the conclusions we argue that this damping rate is directly related to the cusp anomalous dimension of a cusped Wilson loop.)  In section 4, we construct dressed states and we show that, with a sufficiently careful infrared regularization, the dressed states have amplitudes where the leading correction varies linearly in the quark's proper time with precisely the same damping rate as we found for the inclusive formalism.  However a fundamental difference is that the divergences cancel for an incoming dressed quark with any colour and, therefore, the colour of dressed states does not fluctuate and it would seem to be a meaningful quantity.  Section 5 summarizes some conclusions.

  \section{Heavy quark with constant acceleration}

 In this section we will set up some of the technical framework that is needed for our computations.  This is mainly a review of results in reference \cite{Hubeny:2014zna} which are also reviewed  in reference \cite{Semenoff:2018ffq}
 and which  are closely related to previous results using perturbation theory to study the Euclidean circle Wilson loop \cite{Erickson:2000af}
\cite{Semenoff:2001xp}.  We present them here for completeness and to set the context for our use of them in the following sections.
 
 We will consider the following thought experiment.  We begin with ${\mathcal N}=4$ supersymmetric Yang-Mills theory with gauge group U(N+1).  This theory has a single coupling constant, $g_{\rm YM}$ which is freely tuneable as its beta function vanishes.  Moreover, it is believed that there are no phase transitions as this coupling parameter is varied between zero and infinity.  We will study this theory on the Coulomb branch.  For this, we shall assume that one matrix element of one of the scalar fields, say $\left[\Phi^1\right]_{N+1~N+1}$ of the ${\mathcal N}=4$  Yang-Mills theory gets a vacuum expectation value.  The Higgs mechanism then reduces the gauge symmetry from  U(N+1) to U(N)$\times$U(1).  The result is ${\mathcal N}=4$ Yang-Mills theory with gauge group U(N) coupled to a massive Higgs field,  corresponding to the
 fluctuation in the magnitude of the condensate,   $\left<\Phi^1_{N+1~N+1}\right>$, and a massive short super-multiplet of W-bosons.
 That super-multiplet  contains vector, spinor and scalar fields. The $W$-bosons transform in the fundamental representation of the residual U(N) 
 gauge group and they carry a single unit of U(1) charge. The remaining massless ${\mathcal N}=4$ multiplet transform under the adjoint representation of U(N) and they are neutral under the U(1). The mass of the W particles is determined by the value of the condensate.  The magnitude of the condensate is a  tuneable dimensionful parameter which does not get quantum corrections.   Moreover, due to the ${\mathcal N}=4$ supersymmetry, the vacuum energy does not depend on this parameter. It is a supersymmetric modulus.
 
 Our heavy quark will be a massive fundamental representation scalar particle in the W-boson super-multiplet which is put into a state of constant acceleration.  This acceleration could be driven, for example, by a constant external electric field which couples to its conserved U(1) charge. In that case, the mass of the quark must be  large compared to the acceleration in order to suppress  processes which would compete with the one that we are interested in.\footnote{Schwinger pair production of W$^+$-W$^-$ pairs by an constant electric field is an example of such a process. It should be suppressed by the factor $\exp\left( -M^2/\pi E\right)$ where $M$ is the mass of the W particle and $E$ is the magnitude of the electric field.  The proper acceleration is $a=M/E$, so the heavy quark limit  is $\frac{M^2}{E}=\frac{M}{a}>>1$.} 
 
 We inject the heavy quark into the system with an initial  velocity which is in a direction directly opposing the constant acceleration.  Then, we ask the question as to what is the probability amplitude that the particle reemerges at the point where it was injected and after a time which is consistent with the classical motion of such a particle.   The classical particle would emerge at that point, traveling at the same speed in the opposite direction after a time which can be found by solving the classical equation of motion.  In essence, we are asking what is the probability amplitude that the particle follows the classical constant acceleration trajectory.   
  
 The accelerated trajectory which we shall use  is given by the parametric curve 
 \begin{align}
 \tilde x^\mu(\tau)=\left( \frac{1}{a}\sinh a\tau, \frac{1}{a}\cosh a\tau,0,0\right)~,~~-\frac{\tau_P}{2}\leq \tau\leq \frac{\tau_P}{2}
 \label{trajectory}
 \end{align}
 Since $\dot {\tilde x}^\mu(\tau)\dot {\tilde x}_\mu(\tau)= -1$, $\tau$ is the proper time and $\tau_P$ is the total proper time that is experienced by the
 particle during its flight.  Moreover, $\ddot {\tilde x}^\mu(\tau)
 \ddot {\tilde x}_\mu(\tau)=a^2$ and $a$ is the constant proper acceleration.

 Let us first consider the propagation of the particle without coupling to the ${\mathcal N}=4$ fields, that is, in the limit where the 
 Yang-Mills theory coupling constant vanishes, $g_{\rm YM}=0$. 
 The quantum  amplitude is given by the single particle world-line path integral 
 \begin{align}\label{A0}
 {\mathcal A}_{0ab}=\delta_{ab}\int_0^\infty dT\int [dx^\mu(\tau)] e^{iS[x,T]}
 \end{align}
 where $\tau\in[-\tau_P/2,\tau_P/2]$
  and 
 the functions $x^\mu(\tau)$  obey the Dirichlet boundary conditions 
 \begin{align}
 x^\mu(\tau_P/2)=  \tilde x^\mu(\tau_P/2)
=\left( \frac{1}{a}\sinh a\frac{\tau_P}{2}, \frac{1}{a}\cosh a\frac{\tau_P}{2},0,0\right)  \label{bc1}\\
 x^\mu(-\tau_P/2)=  \tilde x^\mu(-\tau_P/2)
=\left(- \frac{1}{a}\sinh a\frac{\tau_P}{2}, \frac{1}{a}\cosh a\frac{\tau_P}{2},0,0\right)  \label{bc2}
\end{align}
The boundary conditions  fix the initial and final positions. 
 Here, since the interaction with ${\mathcal N}=4$ fields is turned off, the colour state of the particle, which is labeled by $a$ and $b$ remains intact during the 
particle's flight. This is the source of $\delta_{ab}$ for the colour indices in equation (\ref{A0}). 
 
 If the acceleration is driven by an electric field,  the world line action would contain a coupling to the electric field
 \begin{align*}
 S[x,T]= \int_{-\frac{\tau_P}{2}}^{\frac{\tau_P}{2}} d\tau\left[-\frac{1}{4T} \dot x^\mu(\tau)\dot x_\mu(\tau)+M^2T +\frac{1}{2}
 E\left( x^0(\tau)\dot x^1(\tau)-x^1(\tau)\dot x^0(\tau)\right)\right]
   \end{align*}
   where $M$ is the mass of the particle, $E$ is the strength of the electric field which we have assumed is in the $x^1$ direction.   
   The last terms, proportional to $E$,  are the line integral of the Abelian vector potential for a constant electric field. 
   The classical equation of motion for this action is solved by the trajectory $\tilde x^\mu(\tau)$ given in equation (\ref{trajectory}), with $a=\frac{E}{M}$
    as well as $\tilde T=\frac{1}{2M}$.  The on-shell action is $S[\tilde x,\tilde T]=-\frac{1}{2}M\tau_P$ and the semi-classical limit of the amplitude
    is a phase,
    \begin{align}\label{amplitude_phase}
    {\mathcal A}_{0ab}= \delta_{ab}e^{-\frac{i}{2}M\tau_P} 
       \end{align}
   which grows linearly in the proper time.  The validity of this approximation is governed by two limits, the heavy quark limit $M>>a$
   and the large time limit $M\tau_P>>1$. 
   
   The heavy quark which we have been describing transforms in the fundamental representation of the U(N) gauge group of the massless ${\mathcal N}=4$ supermultiplet.   If we turn on the interactions of the particle with these massless fields, it couples to the
   U(N)  fields by the insertion of the Wilson line into the world-line path integral, so that the amplitude is given by\footnote{The most general expression for this transition amplitude would include closed loops for $W$-bosons as well as open lines corresponding to $W^+$$W^-$ pair production.    We are assuming  that such loops are suppressed by the large mass limit for the  $W$  particle. Indeed, they would generally be expected to correct our results by positive powers of $\frac{a}{M}$.  Pair production is known to be exponentially suppressed, $\sim e^{-M/\pi a}$. Since $W$'s are bifundamental fields, all of these contributions would also be suppressed by the large N limit which we will sometimes (but not always) take.}    

      \begin{align}\label{A}
 {\mathcal A}_{ab}=\int_0^\infty dT\int [dx^\mu(\tau)] e^{iS[x,T]}\left< W[x]_{ab} \right>
 \end{align}
 where 
 \begin{align}\label{wlo}
 W[x]={\mathcal P}e^{ i\int_{\frac{\tau_P}{2}}^{\frac{\tau_P}{2}}  d\tau[ A_\mu(x(\tau))\dot x^\mu(\tau) + |\dot x(\tau)|\Phi^1(x(\tau)) ] }
 \end{align}
 ${\mathcal P}$ denotes path ordering and 
 the bracket is the vacuum expectation value in the ${\mathcal N}=4$ Yang-Mills theory with U(N) gauge group.  
 
    The expectation value of the Wilson line operator (\ref{wlo}) is not gauge invariant.  A gauge fixing of the Yang-Mills theory after which only the global U(N) remains in intact, must be done before the Wilson line is computed.  
  The expectation value will then depend on which particular  gauge fixing is used.  This gauge dependence would be expected to go away in the limit where $\tau_P\to\infty$ where the world-line path integral describes an $S$-matrix element.  {\bf For this reason, we should only trust and will eventually take the large $\tau_P$ asymptotic limit of the amplitude.   } The only physical information that we will extract is the coefficient of the linear in $\tau_P$ component of the phase, in that large $\tau_P$ limit.

 Using the global U(N) invariance of the state, we can re-write the amplitude as 
   \begin{align}\label{A}
 {\mathcal A}_{ab}=\delta_{ab}\int_0^\infty dT\int [dx^\mu(\tau)] e^{iS[x,T]}\left< \frac{1}{N}{\rm Tr}W[x] \right>
 \end{align}
 The Wilson line contributes to the dynamics of the particle by virtue of its dependence on the trajectory, $x^\mu(\tau)$.  
 This dependence is generally  complicated.  However, we can still say something about it in the heavy quark limit. 
 Consider the effect of the insertion of the Wilson line on the classical equation of motion for the particle, which is corrected to
 \begin{align}
 \frac{1}{2T}\ddot x^1-E\dot x^0+\frac{1}{i}\frac{\delta}{\delta x^1(\tau)}\ln \left< \frac{1}{N}{\rm Tr}W[x] \right>=0 \\
 -\frac{1}{2T}\ddot x^0+E\dot x^1+\frac{1}{i}\frac{\delta}{\delta x^0(\tau)}\ln \left< \frac{1}{N}{\rm Tr}W[x] \right>=0 
 \end{align}
 with the same boundary conditions (\ref{bc1}) and (\ref{bc2}).  
 Due to the spacetime symmetry of the trajectory\footnote{The functional derivative might get contributions from the endpoints
 of the interval,  at $\tau=\pm\tau_P/2$. With the  Dirichlet boundary conditions which we are using, variations of the trajectory vanish at the endpoints, so the equation is correct when $\tau$ is strictly between the endpoints, which is sufficient for our purposes.  More precisely  , $\left. \delta \ln \left< \frac{1}{N}{\rm Tr}W[x] \right>=\int d\tau \delta x^\mu(\tau)\frac{\delta}{\delta x^\mu(\tau)}\ln \left< \frac{1}{N}{\rm Tr}W[x] \right>\right|_{x(\tau)=\tilde x(\tau)}=0$  when $\delta x^\mu(\tau)=0$ at $\tau=\pm\tau_P/2$.}
 \begin{align}
 \left. 
 \frac{\delta}{\delta x^\mu (\tau)}\ln \left< \frac{1}{N}{\rm Tr}W[x] \right>\right|_{x(\tau)=\tilde x(\tau)}=0 ~,~~\tau\in (-\tau_P/2,\tau_P/2)
 \end{align}
 This can also be seen to be a result of the supersymmetry of the trajectory \cite{Semenoff:2004qr}.  The result is that 
 the solution of the classical equation of motion is still  the uniformly accelerated trajectory $\tilde x^\mu(\tau)$ given in (\ref{trajectory})
 with $a=\frac{E}{M}$ and $\tilde T=\frac{1}{2M}$.  Then, in the semi-classical limit,  the amplitude is given by the integrand
 of the functional integral evaluated on the classical trajectory, 
 \begin{align}
 {\mathcal A}_{ab}=\delta_{ab} e^{-i\frac{1}{2}M\tau_P} \left< \frac{1}{N}{\rm Tr}W[\tilde x] \right>
 \end{align}
 We emphasize that the validity of this approximation does not rely on a weak coupling or large N limit.  It relies on the heavy quark limit where the 
 dimensionless parameters which control the approximation is the small parameter $\frac{a}{M}$ and the large parameter $M\tau_P$.

 \subsection{Wilson line for an accelerating trajectory}

 We can easily find  leading terms in the Wilson line expectation value $ \left< \frac{1}{N}{\rm Tr}W[\tilde x] \right>$ at the weak coupling limit by Taylor expanding the
exponential and  using Wick's theorem.   The free-field two-point functions of the vector and scalar fields in the Feynman gauge are
\footnote{Here, we are using the conventions where a covariant derivative of an adjoint representation field is $D_\mu\psi=\partial_\mu\psi-ig_{\rm YM}\left[A_\mu,\psi\right]$ and the Yang-Mills action
is normalized as $S_{\rm YM}=-\frac{1}{2}\int d^4x ~{\rm Tr} F_{\mu\nu}F^{\mu\nu}+\ldots$.}
 \begin{align}\label{gluon}
 \left< A^{\mu}_{ ab}(x) A^\nu_{cd}(y)\right>_0=
 \frac{ \eta^{\mu\nu}\delta_{ad}\delta_{bc} }{8\pi^2(x-y)^2+i\epsilon} \\
 \label{scalar}
 \left< \Phi^{I}_{ ab}(x) \Phi^J_{cd}(y)\right>_0=  
 \frac{ \delta^{IJ}\delta_{ad}\delta_{bc} }{8\pi^2(x-y)^2+i\epsilon} 
 \end{align}
 respectively.

 The essential simplification that allows us to easily do computations is the fact that the combination of 
 vector and scalar field correlation functions between points on the classical trajectory are equal to constants
 \begin{align}
 \dot{\tilde x}^\mu{\dot{{\tilde x}}'}^\nu \left< A^{\mu}_{ ab}(\tilde x) A^\nu_{cd}(\tilde x')\right>_0
+| \dot{\tilde x}||\dot{{\tilde x}}'|\left< \Phi^{1}_{ ab}(\tilde x) \Phi^1_{cd}(\tilde x')\right>_0
=\frac{a^2 \delta_{ad}\delta_{bc} }{16\pi^2}
\label{constant}
\end{align}
A similar phenomenon is well known for  the Euclidean circle Wilson loop \cite{Erickson:2000af}
\cite{Pestun:2007rz} which is related to the present discussion by the fact that the constant acceleration trajectory is
the analytic continuation to Minkowski space of the arc of a Euclidean space circle.

Using the propagators (\ref{gluon}) and (\ref{scalar}), and computing the Wilson loop expectation value to the leading order we find the result for the amplitude
 \begin{align}\label{wl}
{\mathcal A}_{ab}[\tau_P] ~=~e^{-i\frac{1}{2}M\tau_P}\delta_{ab}\left[ 1-\frac{g_{YM}^2N}{32\pi^2}(a\tau_P)^2+\ldots\right]
 \end{align}
 This expression includes the corrections depicted in figure 1.
  \begin{figure}
  \includegraphics[scale=1.2]{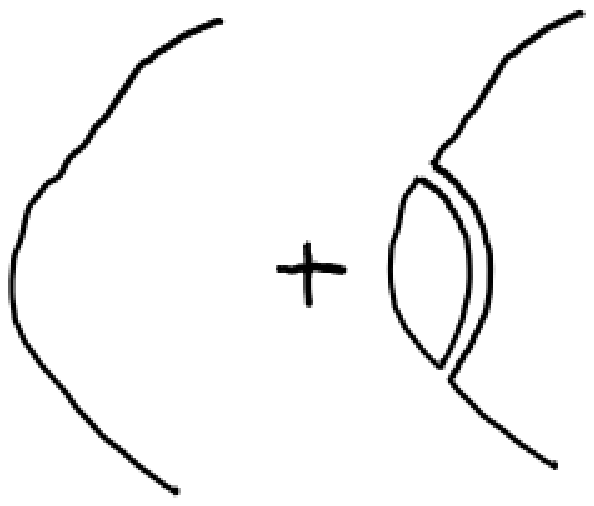}
\begin{caption}{The Feynman diagrams contributing to the corrections to the amplitude that are quoted in equation (\ref{wl})
are depicted in fat graph notation.  The single lines represents the fundamental representation heavy quark and the double line represents the adjoint representation scalar and vector fields.   }
\end{caption}
\label{figure0}\end{figure}
 The leading order correction in equation (\ref{wl}) has the unusual feature that its magnitude grows like the proper time squared.  We interpret this quadratic growth as an infrared divergence.   Indeed, it can easily be seen to come from the dispersive part of the two-point functions, which is the origin of infrared divergences in the usual perturbative computations of Wilson lines.   Moreover, this infrared divergence has the feature that, if we re-sum higher orders, we expect to  obtain a Gaussian damping of the amplitude.  If the gauge group were $U(1)$, for instance, the correction in equation (\ref{wl}) would exponentiate to produce the expression 
 \begin{align}\label{wl_N=1}
  {\mathcal A}(N=1) =   
   e^{-i\frac{1}{2}M\tau_P~-~\frac{g_{\rm YM}^2}{32\pi^2}(a\tau_P)^2}
   \end{align}
    We present some evidence for
    partial exponentiation in   Appendix \ref{ssum_of_ladders}  where we show  that the sum of  ladder diagrams to all orders  changes the leading order result in equation (\ref{wl}) to the result quoted in equation (\ref{sum_of_ladders}) (which we copy here:)
  \begin{align*}
{\mathcal A}_{ab}[\tau_P] ~=~e^{-i\frac{1}{2}M\tau_P}\delta_{ab}~e^{- \frac{g_{\rm YM}^2(a\tau_P)^2}{32\pi^2}}
\frac{1}{N}L^1_{N-1}\left(\frac{g_{\rm YM}^2(a\tau_P)^2}{16\pi^2}\right)
\end{align*}
Here $L^1_{N-1}$ is a polynomial.  
We see that the result still has an exponential  with Gaussian falloff in the proper time of the quark. 
Given that we should take seriously only the large $\tau_P$ asymptotic, we would interpret this as telling us that the amplitude is zero. No matter how heavy the quark and how accurate the semi-classical approximation is expected to be, the probability amplitude for the heavy quark to follow its classical trajectory for a long time is equal to zero.

What went wrong?  It can only be that there are other processes which compete with the one that we have considered, and which take up all of the probability, that is, the other processes are the only likely outcomes when $\tau_P$ is sufficiently large.  When there are massless particles involved, it is clear what these processes are.  They are the emission of soft particles - bremsstrahlung - which should occur when a charged particle accelerates.
The problem with  the above argument is that we have ignored the  amplitudes where the constantly accelerating heavy quark emits bremsstrahlung.   The   probability that an incoming quark emerges as an outgoing quark by itself is zero.  The nonzero probability resides in processes where soft massless particles are produced and also occupy the outgoing state. We will discuss the contributions of soft particle emission in the next section.

 \section{Soft particle emission}

\subsection{Bremsstrahlung and recovering unitarity}

To take into account soft bremsstrahlung, we need to add the probability  that soft particles are produced to the probability of the process that we have already examined, that is, the square of the modulus of the amplitude in equation (\ref{wl}).  For this, it is convenient to describe the 
quantum state of the system using a density matrix.  If the system is in a  pure state $|\psi>$ the  density matrix is  $\rho =|\psi><\psi|$. We could consider  more general mixed states  where $\rho $ is a Hermitian matrix with eigenvalues $P_1,P_2,\ldots,P_n$ and $0\leq P_i\leq 1$ and $\sum_i P_i=1$. In particular ${\rm Tr}\rho=1$.

 \begin{figure}
  \includegraphics[scale=.75]{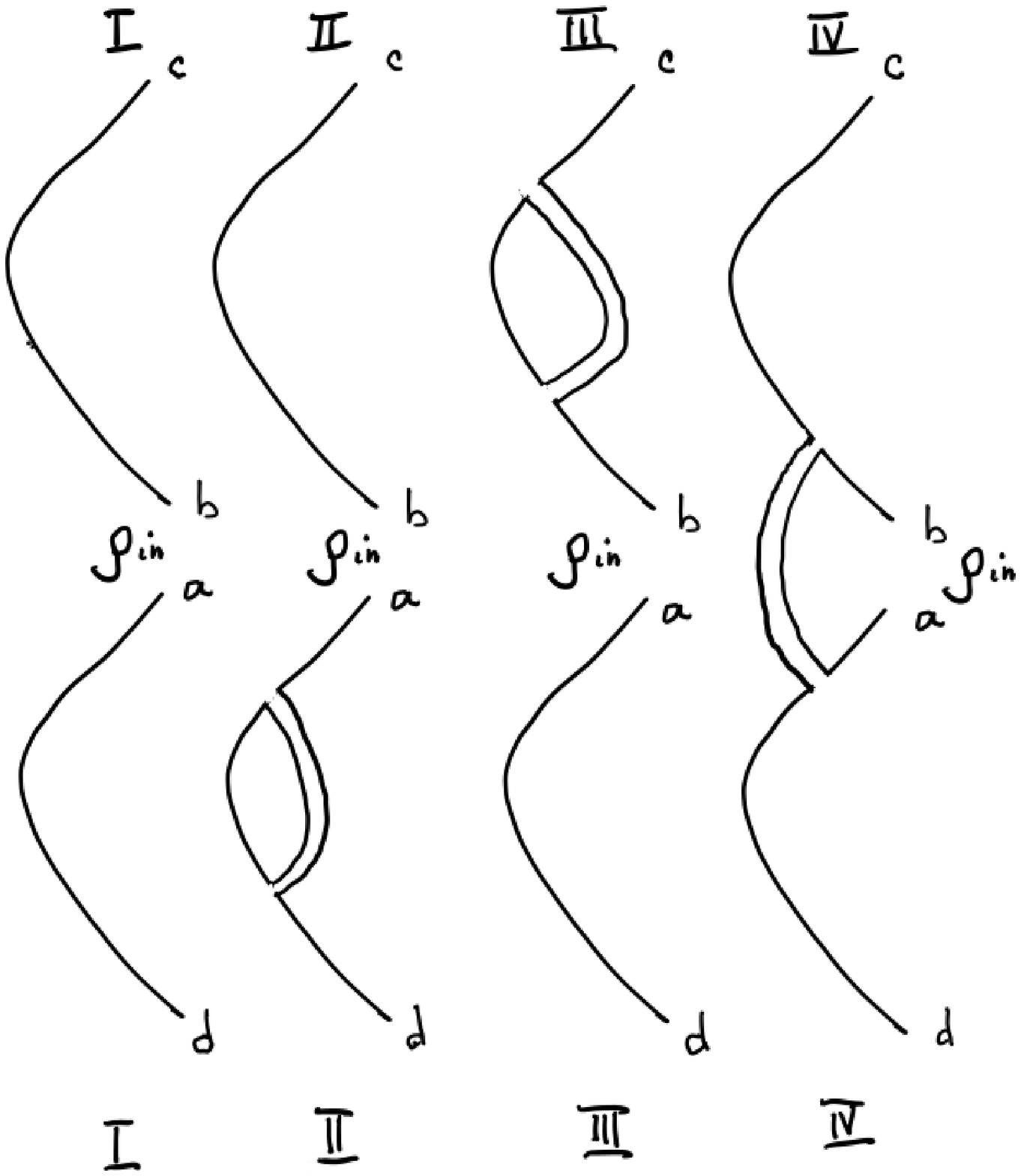}
\begin{caption}{The Feynman diagrams corresponding to the corrections to the evolution operator for the density matrix 
wich take into account the interaction of the heavy quark with massless vector and scalar particles, for order $g_{\rm YM}^2$ are depicted in fat graph notation.  The initial density matrix is at the center and it has colour indices $a,b$.  The reduced out-going density matrix has the indices $d,c$.  Diagram I is the tree-level, diagrams II and III depict the correction to the amplitude coming from the self-interaction of the heavy quark, mediated by exchanging massless vector and scalar fields, and diagram IV  is the correction to the reduced density matrix arising from tracing out the  vector and scalar particles that are produced.  In terms of colour indices, contributions I,II and III have the structure $\sim\delta_{ad}\delta_{bc}$ whereas  IV has the structure $\sim\delta_{ab}\delta_{cd}$. }
\end{caption}
\label{figure1}\end{figure}
 Let us call the incoming
density matrix $\left[\rho_{\rm in}\right]_{ab}$. The amplitude that we are computing evolves the density matrix from the incoming to out-going one
\begin{align}
\left[\rho_{\rm out}\right]_{ab} =\sum_{c,d=1}^N {\mathcal A}_{ac} \left[\rho_{\rm in}\right]_{cd}{\mathcal A}^\dagger_{db}
\end{align} 
which, when we plug in the amplitude that we have computed so far, the result quoted in (\ref{wl}), $\rho_{\rm out}$ becomes
\begin{align*}
\left[\rho_{\rm out}\right]_{ab} = \left[\rho_{\rm in}\right]_{cd}   \delta_{ac}\delta_{bd}\left(1-\frac{g_{\rm YM}^2N}{16\pi^2}a^2\tau_P^2+\ldots\right)
\end{align*}
The fact that we still missing some essential  diagonal elements of the density matrix is seen in 
\begin{align*}
{\rm Tr}\left[\rho_{\rm out}\right] ={\rm Tr} \left[\rho_{\rm in}\right] \left(1-\frac{g_{\rm YM}^2N}{16\pi^2}a^2\tau_P^2+\ldots\right)
\end{align*}
that is, the normalization of the density matrix, which always should have unit trace, decreases with time.

To the leading order in $g_{\rm YM}$, the amplitude for a quark in colour state $a$ to emit  of a single massless vector particle with wave-vector $\vec k$, polarization $e^s_{\mu}(k)$ and with colour indices $cd$ is given by\footnote{The factor of $1/\sqrt{2}$ comes from the unit normalization of the states when the commutation relations for creation and annihilation operators are
as in equations (\ref{comm1}) and (\ref{comm2}). }
\begin{align}\label{qtogluonq}
i g_{\rm YM}  \int_{-\frac{\tau_P}{2}}^{\frac{\tau_P}{2}}d\tau e^s_{\mu}(k)\dot x^\mu(\tau)\frac{ e^{-ik_\mu x^\mu(\tau) } }{\sqrt{(2\pi)^32|\vec k|} }
 \frac{\delta_{ac}\delta_{db} }{\sqrt{2}}
~,~~k^0=|\vec k|
\end{align}
 Here, the heavy quark in the initial state has colour index $a$, the soft vector or scalar field in the final state has colour indices $cd$ and the heavy quark in the final state has colour $b$.  Notice that the colour of the heavy quark is not
 conserved in this process.   It changes from $a$ to $b$.  Of course, this is natural when it emits a soft excitation which itself carries colour. 
 
There is a similar expression for the amplitude for the emission of a massless scalar particle
\begin{align}\label{qtoscalarq}
i g_{\rm YM}  \int_{-\frac{\tau_P}{2}}^{\frac{\tau_P}{2}}d\tau |\dot x (\tau)|\frac{ e^{-ik_\mu x^\mu(\tau) } }{\sqrt{(2\pi)^32|\vec k|} }
\frac{\delta_{ac}\delta_{db} }{\sqrt{2}}
~,~~k^0=|\vec k|\end{align}
We must add these amplitudes to the evolution of the density matrix so that the out-going density matrix has components with one physical massless vector or scalar fields as well as the heavy quark.  Indeed, as we shall see shortly, these contributions  help to maintain the normalization of the density matrix.
We could imagine that we include all of the possible soft particle production compatible with the order in perturbation theory to which we are working, we can then take a trace over the states with soft particles to make a reduced density matrix.  
        
        The reduced density matrix is gotten by taking a partial trace of the outgoing density matrix over
        the states of the vector and scalar field.   We shall want to trace over only the soft modes of these fields.  But, let us
        begin by studying what happens when we trace over all of them.  This will give us a way to check unitarity. 
        
         The trace  over all of the massless vector and scalar degrees of freedom is straightforward and the relevant Feynman diagram is diagram IV   depicted in figure 2.  It produces the following term,
 \begin{align}
 \sum_{c,d=1}^N \left[\rho_{\rm in}\right]_{cd} ~
\delta_{ab}\delta_{cd}\frac{g_{\rm YM}^2}{2}
\int d\tau d\tau' 
\left[\dot {\tilde x} (\tau)\cdot\dot {\tilde x} (\tau') + |\dot {\tilde x}(\tau)||\dot {\tilde x}(\tau')|\right]  \int  d^3k\frac{ e^{ik( \tilde x(\tau)-\tilde x(\tau') )} }{ {(2\pi)^32|\vec k|} }
\label{integral}\end{align}
\begin{align}
= \sum_{c,d=1}^N \left[\rho_{\rm in}\right]_{cd} ~\frac{1}{N}\delta_{ab}\delta_{cd} \frac{g_{\rm YM}^2N}{16\pi^2}a^2\tau_P^2 
\end{align}
in the reduced density matrix.   (This integral can  be gotten from equation (\ref{int201}) in appendix \ref{Appendix0} by first
integrating over $k^\pm$ and then $\tau$ and $\tau'$.)

The reduced density matrix is then, up to order $g_{\rm YM}^2$, 
\begin{align}\label{rdm}
\left[\rho_{{\rm red}}(\tau_P)\right]_{ab} = \sum_{c,d=1}^N\left[\rho_{\rm in}\right]_{cd}  \left\{ \delta_{ac}\delta_{bd} -\frac{g_{\rm YM}^2N}{16\pi^2}a^2\tau_P^2 
  \delta_{ac}\delta_{bd}+\frac{g_{\rm YM}^2}{16\pi^2}a^2\tau_P^2 \delta_{ab}\delta_{cd} +\ldots\right\}
\end{align}
The first term on the right-hand-side is the evolution with no interactions at all. It corresponds to contribution I in figure 2. (The phase in the amplitude (\ref{amplitude_phase}) cancels when we apply it to the evolution of the density matrix.)  The second term is  the leading self-energy correction to the heavy quark.  It corresponds to contributions II and III in figure 2.  The third term is from the trace of the density matrix in the sector  where vector and scalar particles are produced.   It corresponds to contribution IV in figure 2.  Note that the third term has a different structure in colour indices from the other terms.   

Now, we can confirm that the trace of the reduced density matrix (and therefore also the trace of the outgoing density matrix) is equal to the trace of the incoming density matrix,
\begin{align}
\sum_a\left[\rho_{{\rm red}}(\tau_P)\right]_{aa} =\sum_a\left[\rho_{{\rm in}}(\tau_P)\right]_{aa} 
\end{align}
  This means that we have taken into account every possible process that can occur in the leading orders of perturbation theory
and in the limit where the quark is heavy -- we have recovered unitarity. 

The infrared divergent terms, those proportional to $(a\tau_P)^2$ remain in the expression (\ref{rdm}).  This means that only certain incoming density matrices, or certain questions involving outgoing density matrices are not infected  with infrared divergences.  Explicitly, we can rewrite equation (\ref{rdm}) as
\begin{align}\label{rdm1}
\left[\rho_{{\rm red}}(\tau_P)\right]_{ab} = \left[\rho_{{\rm in}}  \right]_{ab} -\frac{g_{\rm YM}^2N}{16\pi^2}a^2\tau_P^2 
\left( \left[ \rho_{{\rm in}}\right]_{ab}-\frac{1}{N}\delta_{ab} \right)+\ldots
\end{align} 
where we have used ${\rm Tr}\left[\rho_{\rm in}\right]=1$. 
We see that the divergent terms are absent only when
\begin{enumerate}
\item{}  the incoming density matrix
is $ \left[\rho_{{\rm in}}\right]_{ab} = \frac{1}{N}\delta_{ab} $, that is, the initial state is  a  singlet mixed state of colour where all of the colours have the same probability or\\
\item{}the observables whose expectation values we calculate using  the final density matrix  are averaged over colour states
 $<{\mathcal O}>={\rm Tr}\left[ {\mathcal O}\rho_{{\rm red}}(\tau_P) \right] $ and 
 ${\mathcal O}_{ab}\sim \delta_{ab}\cdot$(independent of $a$ and $b$).
\end{enumerate}

\subsection{Introducing a detector resolution}
  
It is usual to cancel infrared divergences by considering inclusive probabilities for soft particle production, rather than inclusive of all of the inelastic behaviour as we have done so far in the discussion above.  This means that we should trace the outgoing density matrix over only the soft vector and scalar particles which are emitted.   The trace then would have an upper cutoff on wave-vectors which is intended to emulate a detector resolution, where the hypothesis is that we must always compute inclusive probabilities where we include the production of any soft particles which would escape detection.   We will do this shortly. 
 However, as a first pass, we could simply assume that the ``detector''  is switched on and it can only detect massless particles which are emitted in the time interval $[-\hat x^0/2,\hat x^0/2]$ where 
  \begin{align}
  \hat x^0 = \frac{2}{a}\sinh a\frac{\sigma_P}{2} ~,~~\sigma_P<<\tau_P
  \end{align}
  That means that we should trace over massless particles which are produced in the proper time intervals where the detector is switched off, that is $[-\frac{\tau_P}{2},-\frac{\sigma_P}{2}]$ and $[\frac{\sigma_P}{2},\frac{\tau_P}{2}]$. 
 Doing so yields the reduced density matrix 
  \begin{align}
\left[\hat \rho_{{\rm red}}\right]_{ab} &= \left[\rho_{{\rm in}}  \right]_{ab} -\frac{g_{\rm YM}^2N}{16\pi^2}a^2\tau_P^2 
\left[ \rho_{{\rm in}}\right]_{ab}
+\frac{1}{N}\delta_{ab} \frac{g_{\rm YM}^2N}{16\pi^2}a^2(\tau_P-\sigma_P)^2\ldots \nonumber  \\
 &= 
\left[\left[ \rho_{\rm in}\right]_{ab}-\frac{1}{N}\delta_{ab}\right]\left[1 -\frac{g_{\rm YM}^2N}{16\pi^2}(a\tau_P)^2+\ldots \right]
\nonumber \\ &~~~~~~~~~~~~~~~~~~~~~~~~
+\frac{1}{N}\delta_{ab}\left[1- \frac{g_{\rm YM}^2N}{8\pi^2}(a\sigma_P)(a\tau_P)+\ldots \right]
\label{rdm11}
\end{align}
Here, we have assumed that $\tau_P>>\sigma_P$ and we have dropped terms that 
are less than linear in $\tau_P$.  
 This is similar to the previous result, where we traced out all of the particles produced, in that the terms which are quadratic in $\tau_P$ will be absent only when the initial density matrix satisfies $\left[ \rho_{\rm in}\right]_{ab}-\frac{1}{N}\delta_{ab}=0$ or only when we use this density matrix to take the expectation value of a colour singlet operator.
 However, there remains a linear term in the total proper time.  This is due to the fact that the possibility of the production of hard vector and scalar particles has been left out of our reduced density matrix.  The rate at which they are being produced is \begin{align}
\Gamma = \frac{g_{\rm YM}^2N}{8\pi^2}(a\sigma_P)a
\end{align}
or, in terms of the total rest frame time during which the detector is active, $\hat x^0$, 
\begin{align}
\Gamma =  \frac{g_{\rm YM}^2N}{4\pi^2}a\ln(a\hat x^0)
\label{damping_rate}
\end{align}
This damping rate depends on the detector resolution in the sense that it depends logarithmically on
the time $\hat x^0$ that the detector is switched on, in units of the acceleration.

The normalization of the density matrix does not remain unity, but decreases with time because it is not the full density matrix.  We have left out the competing processes consisting of the production of hard vector and scalar particles. If those processes were included, as they were in the discussion of the previous subsection, we would find unit trace. The probability of detecting the outgoing quark without hard particle production decreases with time simply because the probability of having produced some hard vector and scalar particles increases with time.
 
 \subsection{Introducing a fundamental infrared cutoff}
 
 A more conventional alternative to the above scheme for introducing a detector resolution would be to impose wave-vector cutoffs in the integrals over the wave vectors of  massless particles that are encountered in the computations of the corrections to the out-going density matrix.  This would be a fundamental cutoff which we shall call $m$, which will be needed in order to define the Feynman  integrals in the first place and a detector resolution which we shall call $\Lambda$.  We will assume a large hierarchy of these scales, 
\begin{align*}
m<<\Lambda<<a<<M
\end{align*}
 with $M$ the heavy quark mass.  ``Soft particles'' will be those with energies between the fundamental cutoff $m$ and the detector resolution, $\Lambda$.  ``Hard particles'' will be those with energies greater than $\Lambda$. 

Once the fundamental cutoff $m$ is introduced, and after extracting a factor of overall proper time $\tau_P$, we can safely take  the limit $\tau_P\to \infty$ and do the remaining proper time and momentum integrals explicitly. The procedure is summarized in 
 appendix \ref{Appendix0}.    
 The result for the out-going reduced density matrix is 
  \begin{align} 
\left[\hat \rho_{{\rm red}}\right]_{ab} 
&=\frac{\delta_{ab}}{N} \left(1-\frac{g_{\rm YM}^2N}{4\pi^2}\left[\ln\frac{a}{\Lambda}\right](a\tau_P) \right) 
\nonumber \\ &+
\left[\left[ \rho_{\rm in}\right]_{ab}-\frac{1}{N}\delta_{ab}\right]\left(1-\frac{g_{\rm YM}^2N}{4\pi^2}\left[\ln\frac{a}{m} \right](a\tau_P)\right)
+\ldots 
\label{rdm2}
 \end{align}

 Note that the quadratic dependence on $\tau_P$ is now absent.  It has been replaced by linear terms in $\tau_P$ times logarithms of the fundamental  infrared cutoff $m$.  
Again, as we see clearly in the expression (\ref{rdm2}), the infrared divergent terms which are logarithms of the fundamental cutoff  cancel only when the incoming density matrix is proportional to the unit matrix in colour indices, 
\begin{align*}
   \left[\rho_{{\rm in}}  \right]_{ab}=\frac{1}{N}\delta_{ab}
 \end{align*}
 The decay rate of the state,
 per unit of the heavy quark's proper time is 
 \begin{align}\label{decay_rate}
 \Gamma = \frac{g_{\rm YM}^2N}{4\pi^2}a \ln\frac{a}{\Lambda} 
 \end{align}
   Note the coincidence of the decay rates that we have computed in equations (\ref{damping_rate}) and (\ref{decay_rate}). 
   It associates the detector resolution $\Lambda$ with the inverse of the length of the rest frame
   time during which the detector was operating, $(\hat x^0)^{-1}$.

 \section{  Dressed States}
 
 There is an alternative way of dealing with the infrared which uses dressed states.   In that approach, instead of computing transition amplitudes for single particle states,  as we have done in the previous section, we dress the incoming and outgoing states with clouds of soft particles.  The dressed states should contain the soft particles that are produced by the interactions
 during the process.   Then we compute the amplitude for a transition between two dressed states by taking matrix elements of the $S$-matrix. 
 
 To the order in perturbation theory that we are working, ( order $g_{\rm YM}^2$ for diagonal elements, order $g_{\rm YM}$ for off-diagonal elements),  and in the single heavy quark sector, the $S$-matrix is 
\begin{align}
S_{ab}&=e^{-i\frac{M}{2}\tau_P}\delta_{ab}\left(1 -\frac{g_{\rm YM}^2N}{8\pi^2}(a\tau_P)\ln\frac{a}{m}\right) +
\nonumber  \\
&+
ig_{\rm YM} \int_m\frac{d^3k}{\sqrt{(2\pi)^32|k|}} \int_{-\tau_P/2}^{\tau_P/2}d\tau e^{ikx(\tau)}  
\left[\dot x^\mu(\tau)a_{\mu ab} (k)+| \dot x(\tau)|a_{  ab}^{1 }(k) \right]
\nonumber \\
&+ig_{\rm YM} \int_m\frac{d^3k}{\sqrt{(2\pi)^32|k|}} \int_{-\tau_P/2}^{\tau_P/2}d\tau e^{-ikx(\tau)}  
\left[\dot x^\mu(\tau)a_{\mu ab}^\dagger (k)+| \dot x(\tau)|a_{  ab}^{1\dagger }(k) \right] \nonumber  \\
&+\ldots
\label{S_matrix}\end{align}
Matrix elements of this $S$-matrix produce the amplitudes in equations (\ref{qtogluonq}) and (\ref{qtoscalarq}) that we used in
the previous sections  . 
Here, the momentum integrals have a fundamental infrared cutoff $m$.   Note that, like the Wilson line, the $S$ matrix also contains both the vector and scalar fields in a particular combination which allows the $S$-matrix to commute with half of the supersymmetry transformations.  The normalization of the creation and annihilation operators are such that their commutation relations are
\begin{align}
 \left[a^\mu_{ab}(k), a^\dagger_{\nu cd}(k')\right]=\delta^\mu_{~\nu}\delta^3(\vec k-\vec k') \frac{\delta_{ad}\delta_{bc}}{2}
\label{comm1} \\
 \left[a^I_{ab}(k), a^{J\dagger}_{ cd}(k')\right]=\delta^{IJ}\delta^3(\vec k-\vec k') \frac{\delta_{ad}\delta_{bc}}{2}
\label{comm2}
 \end{align}

The dressed state contains   soft massless particles  as well as the heavy quark. 
It must contain only data referring to the final state. Here, we take the dressing as being very similar to that which is used 
in Abelian gauge theory, containing the  four-velocity of the particle.   The dressed final state is 
\begin{align}
&|a>>_{\rm f}\equiv \\ &|a>+g_{\rm YM} \int_m^\Lambda\frac{d^3k}{\sqrt{(2\pi)^32|k|}} \frac{\left[
\dot x^\mu(\tau_P/2)a_{\mu ab}^\dagger(k)+| \dot x(\tau_P/2)|a_{ ab}^{1 \dagger}(k)\right] }
{k_\mu\dot x^\mu(\tau_P/2)}~|b>\\&+\ldots
\label{dressed_out}\end{align}
and the dressed initial state is
\begin{align}
&|a>>_{\rm i}\equiv  \\ & |a>+ g_{\rm YM}\int_m^\Lambda\frac{d^3k}{\sqrt{(2\pi)^32|k|}} \frac{\left[
\dot x^\mu(-\tau_P/2)a_{\mu ab}^\dagger(k)+| \dot x(-\tau_P/2)|a_{ ab}^{1 \dagger}(k)\right] }
{k_\mu\dot x^\mu(-\tau_P/2)}~|b>\nonumber \\ &+\ldots
\label{dressed_in}
\end{align}
Note that, because of the identity $|\dot x(\pm\tau_P/2)|^2+\dot x^\mu(\pm\tau_P/2) \dot x_\mu(\pm\tau_P/2)=0$, the dressed state is normalized, up to order $g_{\rm YM}^2$.  
The question that we are asking is about the amplitude that, when system begins in
state $|a>>_{\rm i}$, it appears in the state $|b>>_{\rm f}$ after a time  $\hat x^0=\frac{2}{a}\sinh (a\tau_P/2)$,
that is,  
\begin{align}\label{dressed_amplitude}
\hat{\mathcal A}_{ab}= ~_{\rm f}<<b|S|a>>_{\rm i}
\end{align}
which we now have the tools to compute  
up to and including order $g_{\rm YM}^2$ in perturbation theory. 

   Putting  together equations (\ref{S_matrix}) to (\ref{dressed_amplitude}), we find
   \begin{align}
   &\hat{\mathcal A}_{ab}=\delta_{ab}- 
 \delta_{ab} \frac{g_{\rm YM}^2N}{2} \times\bigg\{ (I)+(II)+(III)+(IV)\biggr\} 
 +\ldots
 \nonumber \\
  &I= \int_m^\Lambda\frac{d^3k}{ (2\pi)^32|k^0|}    
 \frac{  \dot{\tilde x}^\mu(\tau_P/2) \dot{\tilde x}_{\mu}(-\tau_P/2) +| \dot{\tilde x}(\tau_P/2) || \dot{\tilde x}(-\tau_P/2)| } { [\dot{\tilde x}^\mu(-\tau_P/2) k_\mu][\dot{\tilde x}^\mu(\tau_P/2) k_\mu ] } 
   \nonumber\\
 & II=i  \int_m^\Lambda\frac{d^3k}{(2\pi)^32|k^0|} 
   \int_{-\tau_P/2}^{\tau_P/2}  d\tau e^{ ik_\mu \tilde x^\mu(\tau ) }   \frac{  \left[ \dot{\tilde x}^\mu(\tau ) \dot{\tilde x}_{\mu}(-\tau_P/2) +| \dot{\tilde x}(\tau ) || \dot{\tilde x}(-\tau_P/2) |\right] } {[ \dot{\tilde x}^\mu(-\tau_P/2) k_\mu ]} 
  \nonumber \\
 &III= i
 \int_m^\Lambda\frac{d^3k}{(2\pi)^32|k^0|} 
i  \int_{-\tau_P/2}^{\tau_P/2}  d\tau  e^{-ik\tilde x(\tau) }   \frac{  \left[ \dot{\tilde x}^\mu(\tau_P/2) \dot{\tilde x}_{\mu}(\tau') +| \dot{\tilde x}(\tau_P/2) || \dot{\tilde x}(\tau') |\right] } {[  \dot{\tilde x}^\mu(\tau_P/2) k_\mu ]} \nonumber  \\
 &IV=    -\frac{1}{2} \int_m^\infty\frac{d^3k}{(2\pi)^32|k^0|} 
 \int_{-\tau_P/2}^{\tau_P/2}  d\tau    d\tau' 
 e^ {ik(  x(\tau )-  x(\tau')) }    \left[  \dot {\tilde x}^\mu(\tau) \dot {\tilde x}_\mu(\tau') +|\dot{\tilde x}(\tau)||\dot{\tilde x}(\tau')|\right] 
\end{align}
In each of these integrals,  $I,II,III$ and $IV$, we will extract an overall factor of $\tau_P$ and 
then take the large $\tau_P$ limit of the remaining expressions. Each of 
 the resulting integrals  contain integrations over the momenta $\vec k$ which are infrared divergent.   In order to define those momentum integrations, we will introduce a fundamental infrared regulator, $m$, in each of the integrals by modifying the momentum integral.  We will replace photon mass shell condition $k^0=|\vec k|$ by the massive dispersion relation $k^0=\sqrt{\vec k^2+m^2}$.
 $$\int_m^\infty\frac{d^3k}{(2\pi)^32|k^0|} \ldots \equiv\lim_{m\to 0} \int \frac{d^4k}{(2\pi)^3}\delta(k^\mu k_\mu+m^2)\theta(k^0)\ldots
 $$
 Then, the infrared divergence is manifest in a factor of $\ln(m)$ in each integral.  
  We shall see shortly that these logarithms cancel when the above four integrals are added together. 
Beyond this, the integrals $I$, $II$ and $III$ have an upper cutoff, so that their integrals are also restricted to the region with frequencies less than $\Lambda$.  We achieve this, and preserve as much Lorentz symmetry as possible, by defining the cutoff integral as
$$\int_m^\Lambda\frac{d^3k}{(2\pi)^32|k^0|} \ldots \equiv
\lim_{m\to 0} \int \frac{d^4k}{(2\pi)^3}[\delta(k^\mu k_\mu+m^2)-\delta(k^\mu k_\mu+\Lambda^2) ] \theta(k^0)\ldots
 $$
where $\Lambda$ is the detector resolution. After using the delta functions for the $k^0$-integral, the integrand has support in the region $ |\vec k|\lesssim\Lambda$.  This restricts the soft particle content of the dressing to particles with frequencies and wave-numbers smaller than the detector resolution.  We will work in the limit where both
cutoffs are small compared to the acceleration and where the fundamental cutoff is much smaller than the detector resolution,
$m<<\Lambda <<a<<M$.  The integrals are elementary and some details about how they are done are given in appendices \ref{AppendixI}-\ref{AppendixIV}. The result for the amplitude is 

\begin{align}
   &\hat{\mathcal A}_{ab}=\delta_{ab}\left( 1 - 
\frac{g_{\rm YM}^2N}{8\pi^2}(a \tau_P) \ln\frac{a}{\Lambda}+\ldots\right)
 \end{align}
 
 It is infrared finite and it depends on the detector resolution $\Lambda$.  What is more, the decay rate, 
 $\Gamma=\frac{g_{\rm YM}^2N}{4\pi^2}a   \ln\frac{a}{\Lambda}$ is identical to the one that we found for the  colour neutral incoming density matrix in the inclusive formalism, equation (\ref{decay_rate}).  For that particular state, the  inclusive and the dressed formalisms give identical answers.   But, the main difference of the two is that, in the dressed formalism, colour is conserved.  Whatever the incoming state, the colour indices are simply copied to the outgoing state.  To this order in perturbation theory, colour still seems to be a good physical attribute of the dressed heavy quark.
 
 \section{Conclusions}
 
 In conclusion, we have examined a Gedanken experiment where we asked the question as to what the probability is that a  constantly accelerated heavy quark follows its classical trajectory.  The answer, due to infrared divergences,  was zero.  Then, we took the fact that the quark must emit soft gluons and scalar particles into account and we found that the quadratic terms in proper time which we attributed to infrared divergences cancel when the incoming quark is in a mixed state which is a flat superposition of colour states, $\sim \frac{1}{N}\sum_a |a><a|$. The quadratic dependence on proper time is replaced by a linear divergence and we interpret the coefficient of the linear terms as the inclusive rate at which hard particles are produced,   $\Gamma=\frac{g_{\rm YM}^2N}{4\pi^2}a   \ln\frac{a}{\Lambda}$ where $a$ is the proper acceleration and $\Lambda$ is the detector resolution.  
 
 The coefficient of the acceleration times the logarithm  in $\Gamma$ is identical to the cusp anomalous dimension for the ${\mathcal N}=4$ Wilson loop.  This is not a coincidence. If we parameterize the trajectory using the rest-frame time, $\kappa$, with $-\kappa_{\rm max}/2<\kappa<\kappa_{\rm max}/2$, 
 so that
 $$
 \tilde x^\mu(\kappa) = \frac{1}{a} \left( a\kappa, \sqrt{1+a^2\kappa^2},0,0\right)~\to~\left(\kappa, |\kappa|,0,0\right) ~{\rm as}~ a\to\infty
 $$
 we see that in the large acceleration limit, the trajectory is two null curves with a cusp due to an infinite instantaneous acceleration at $\kappa=0$. In this case, the parameter $a\tau_P =2 \sinh^{-1}a\kappa_{\rm max}/2$ is simply the diverging cusp angle and the logarithmic terms are the usual infrared logs that go with the cusp anomalous dimension.

 We also obtained the same value of the damping constant $\Gamma$ in the dressed state approach.    However, the inclusive approach could only be applied to incoming quarks in colourless mixed states.   For dressed quarks, this was not the case, the in-going to out-going colour index of the dressed quark was conserved.

 We have done explicit computations only to the leading orders of perturbation theory.  It would be interesting to understand whether the phenomena that we have discussed here also appear at  higher orders in perturbation theory.  For example, it would be nice to understand whether collinear divergences due to corrections of the processes where soft gluons are emitted pose a fundamental problem.  Also, it would be interesting to understand whether the matrix model computation is accurate in that whether supersymmetry can be relied on to cancel the corrections to it from Feynman diagrams with internal vertices, in the same way as occurred for the Euclidean circle Wilson loop.  These open questions remain the subject of ongoing work.
  
 It would be interesting to understand whether our results have implications for gluon scattering amplitudes in quantum chromodynamics.  In that theory, the infrared problem is solved dynamically, by confinement.  However, there is a high energy
 regime where perturbation theory is accurate and an interesting question deserving further study is whether our information theoretic considerations possibly have observable effects there. 
 
 Another interesting question is whether what we have described here has a strong coupling dual in string theory.  Constantly accelerated quarks have been studied at strong coupling Yang-Mills theory \cite{Hubeny:2014zna} \cite{Chernicoff:2010yv}-\cite{Hubeny:2014kma} and it was observed that amplitudes for bremsstrahlung are suppressed at large $N$.  However, tracing out soft bremsstrahlung would also involve factors of $N$ and it would be interesting to see whether similar phenomena to what we have considered here would be visible at strong coupling. 

\begin{appendix}

\section{Appendix:  The matrix model sums ladder diagrams}
\label{ssum_of_ladders}

\subsection{One-point function}
We observed in equation (\ref{constant}) that the sum of the vector and scalar two-point functions with endpoints on the contour is a constant.  This is attributable to the supersymmetric nature of the Wilson line, in that it commutes with half of the super symmetries and conformal super symmetries of the ${\mathcal N}=4$ Yang-Mills theory.  We will not discuss the details here.  A brief discussion can be found in reference \cite{Hubeny:2014zna}.   With the effective propagator a constant, we can sum a large class of Feynman diagrams which contribute to the Wilson line, those diagrams which have no internal vertices, by simply solving the combinatorics of the Lie algebra indices on the propagators and vertices.  This can be done using a matrix model,
\begin{align}
\left< W_{ab}[\tilde x]\right> ~=~\frac{1}{N}\delta_{ab}~ \frac{\int [dM] e^{-8\pi^2{\rm Tr}M^2}  [{\rm Tr} e^{ig_{\rm YM}a\tau_P  M}]}{\int [dM]    e^{-8\pi^2{\rm Tr}M^2} }
\label{matrix_model}
\end{align}
where $M$ is a Hermitian $N\times N$ matrix.  This matrix integral can readily be found exactly by using the method of orthogonal polynomials, for example.  The result is 
 \begin{align}\label{sum_of_ladders}
\left< W_{ab}[\tilde x]\right> ~=~\delta_{ab}~e^{- \frac{g_{\rm YM}^2(a\tau_P)^2}{32\pi^2}}
\frac{1}{N}L^1_{N-1}\left(\frac{g_{\rm YM}^2(a\tau_P)^2}{16\pi^2}\right)
\end{align}
where $L^k_n(x)$ is the associated Laguerre polynomial
\begin{align}
L^k_n(x)=\frac{ e^x x^{-k}}{n!}\frac{d^n}{dx^n}\left( e^{-x}x^{n+k}\right)
=\sum_{m=0}^n x^m\frac{(n+k)!}{(n-m)!(k+m)!m!}
\end{align}
and 
\begin{align}
L^1_{N-1}(x)=\frac{ e^x x^{-k}}{n!}\frac{d^n}{dx^n}\left( e^{-x}x^{n+k}\right)
=\sum_{m=0}^{N-1} (-x)^m\frac{(N)!}{(N-1-m)!(1+m)!m!} \nonumber \\
=N-\frac{N(N-1)}{2}x+\ldots
\end{align}
We can see that this result reproduces our perturbative computation by expanding to first order in $g_{\rm YM}^2$. 
We can also see that it reproduces the exponentiation of the Abelian limit, $N=1$ where $L_0^1=1$.

 \subsection{Two-point function}
 
 The sum of all ladder-like Feynman diagrams contributing to the evolution operator, plus the result of tracing over all of the soft particle emission, the leading orders of which  are displayed in figure 2 is 
 given by the matrix model correlation function
  \begin{align}\label{ssssum_of_ladders}
\left< W_{ad}[\tilde x]W^\dagger_{bc}[\tilde x]\right> ~=~\frac{\int [dM] e^{-8\pi^2{\rm Tr}M^2}  [  e^{ig_{\rm YM}a\tau_P  M}]_{ad} [ e^{-ig_{\rm YM}a\tau_P  M}]_{bc} }{\int [dM]    e^{-8\pi^2{\rm Tr}M^2}  }
 \end{align}
 One can check by expanding to second order in $g_{\rm YM}$ that the expansion contains the four terms which reproduce the contributions I,I,III and IV in figure 2.  The sum of all such diagrams of a ladder-like nature, that is, diagrams which only have propagators which begin and end on the contours is given by the matrix integral in equation (\ref{ssssum_of_ladders}). Moreover, the integral is elementary.   $U(N)$ symmetry of the measure tells us that it has the form
 \begin{align*}
&\left< W_{ad}[\tilde x]W^\dagger_{cb}[\tilde x]\right> ~\\ &=~\frac{1}{N^2-1}\left(\delta_{ad}\delta_{bc}-\frac{1}{N}\delta_{ab}\delta_{cd}\right) 
\frac{\int [dM] e^{-8\pi^2{\rm Tr}M^2}  [{\rm Tr} e^{ig_{\rm YM}a\tau_P  M}]  [{\rm Tr} e^{-ig_{\rm YM}a\tau_P  M}]  }{\int [dM]    e^{-8\pi^2{\rm Tr}M^2}  }
\\ &+\frac{N}{N^2-1}\left(\delta_{ab}\delta_{cd }-\frac{1}{N}\delta_{ad}\delta_{bc}\right)
 \end{align*}
 The computation of this two-point function is again an exercise in the application of orthogonal polynomials to matrix integrals.  The generic result is
 \begin{align}
 &\frac{  \int[dM] e^{-\frac{1}{2}{\rm Tr}M^2} [{\rm Tr}e^{i\alpha M} ] [ {\rm Tr}e^{-i\beta M} ]  } { \int [dM] e^{- \frac{1}{2} {\rm Tr} M^2} }=\nonumber \\ & =
 e^{-(\alpha-\beta)^2/2}L^1_{N-1}((\alpha-\beta)^2)+ e^{-(\alpha^2+\beta^2)/2}L^1_{N-1}(\alpha^2)L^1_{N-1}(\beta^2) \nonumber \\ &
 -Ne^{-(\alpha^2+\beta^2)/2 }\frac{L_{N-1}(\alpha^2)L^0_N(\beta^2)-L_N^0(\alpha^2)L_{N-1}^0(\beta^2) }{\alpha^2-\beta^2}
  \nonumber \\& -2e^{-(\alpha^2+\beta^2)/2} \sum_{k=1}^{N-1}(\alpha\beta)^k\frac{(N-k)!}{(N-1)!}
 \frac{L^k_{N-1-k}(\alpha^2)L^k_{N-k}(\beta^2) - L^k_{N-k}(\alpha^2)L^k_{N-1-k}(\beta^2) }{ \alpha^2-\beta^2 }
 \end{align}
   and the special case that we need is
    \begin{align}
 &\frac{  \int[dM] e^{-\frac{1}{2}{\rm Tr}M^2} [{\rm Tr}e^{i\alpha M} ] [ {\rm Tr}e^{-i\alpha M} ]  } { \int [dM] e^{- \frac{1}{2} {\rm Tr} M^2} }=\nonumber \\ & = N 
 -Ne^{-\alpha^2 } [  L^0_{N-1}(\alpha^2)L^1_{N-1}(\alpha^2)-L_N^0(\alpha^2)L_{N-2}^0(\alpha^2)  ]
  \nonumber \\& -2e^{-\alpha^2 } \sum_{k=1}^{N-1}(\alpha)^{2k}\frac{(N-k)!}{(N-1)!}
[ L^k_{N-1-k}(\alpha^2)L^{k+1}_{N-k}(\alpha^2) - L^k_{N-k}(\alpha^2)L^{k+1}_{N-2-k}(\alpha^2) ]
 \end{align}
 and, applied to our case,
  \begin{align*}
&\left< W_{ad}[\tilde x]W^\dagger_{cb}[\tilde x]\right> = \frac{1}{N + 1}\left(\delta_{ab}\delta_{cd }+\delta_{ad}\delta_{bc}\right)~\\ &=~\frac{1}{N^2-1}\left(\delta_{ad}\delta_{bc}-\frac{1}{N}\delta_{ab}\delta_{cd}\right)    
 e^{\frac{-g_{\rm YM}^2a^2\tau_P^2}{16\pi^2}}\left[N^2-N+\ldots\right] 
  \end{align*}
  which gives a late time reduced density matrix
 \begin{align*}
 [\rho_{\rm red}]_{ab}=  \frac{1}{N + 1}\left(\delta_{ab}+ [\rho_{\rm red}]_{ab}\right)+~{\rm exponentially~suppressed}
   \end{align*}
   This results leaves out all of the diagrams which contain any of the Yang-Mills interaction vertices and we have no right to expect that it is accurate in an expansion beyond the order $g_{\rm YM}^2$ which we have computed explicitly.   Whether this observation can be used to obtain a more general result is clearly an interesting open problem.
   
\section{Appendix: Infrared cutoff integration}
\label{Appendix0}
  The integral that we are interested in is 
  \begin{align}\label{int200}
&\frac{g_{\rm YM}^2}{2}\int d\tau d\tau' 
\left[\dot {\tilde x} (\tau)\cdot\dot {\tilde x} (\tau') + |\dot {\tilde x}(\tau)||\dot {\tilde x}(\tau')|\right]  \int_m^\Lambda  d^3k\frac{ e^{ik( \tilde x(\tau)-\tilde x(\tau') )} }{ {(2\pi)^32|\vec k|} } 
\end{align}
where we will impose a lower cutoff $m$ and an upper cutoff $\Lambda$ on the magnitude of $\vec k$, the momentum
of massless particles. 
This integral can be put in a nicer form by noticing that the trajectory of the particle only involves the $x^0$ and $x^1$ directions. 
This allows us to replace the momentum space integral   $$ \int \frac{ d^3k  }{ {(2\pi)^32|\vec k|} } f(k^0,k^1)=
  \int  \frac{ d^4k  }{ (2\pi)^3 } \delta(k_\mu k^\mu)\theta(k^0)f(k^0,k^1)$$ $$=
   \int_0^\infty\frac{dk^+}{2\pi}\int_0^\infty \frac{dk^-}{2\pi}f(k^0,k^1)$$ where we have introduced the light-cone coordinates 
   $$k^\pm = \tfrac{1}{2}(k^0\pm k^1)$$ 
   Moreover 
$$\dot {\tilde x} (\tau)\cdot\dot {\tilde x} (\tau') + |\dot {\tilde x}(\tau)||\dot {\tilde x}(\tau')|=\frac{1}{2} \left[ 2-\tfrac{e^{a\tau}}{e^{a\tau'}}-\tfrac{e^{a\tau'}}{e^{a\tau}}\right]$$ and
  $$
 e^{ik( \tilde x(\tau)-\tilde x(\tau') )}=e^{\tfrac{i}{a} [k^+ e^{-a\tau}-k^-  e^{a\tau}] }
 e^{-\tfrac{i}{a}[ k^+ e^{-a\tau'}-k^-e^{a\tau'}] }
 $$ 
  We thus present the integral as
   \begin{align}
&\frac{g_{\rm YM}^2}{4} \int_{-\tau_P/2}^{\tau_P/2}  d\tau 
 d\tau'
 \left[ 2-\tfrac{e^{a\tau}}{e^{a\tau'}}-\tfrac{e^{a\tau'}}{e^{a\tau}}\right] \cdot \nonumber \\ &\cdot
\int_0^\infty\frac{dk^+dk^-}{4\pi^2}e^{\tfrac{i}{a} [k^+ e^{-a\tau}-k^-  e^{a\tau}] }
 e^{-i\tfrac{i}{a}[ k^+ e^{-a\tau'}-k^-e^{a\tau'}] }\theta(k^+k^--m^2)\theta(\Lambda^2-k^+k^-)
\label{int201} \\&
=\frac{g_{\rm YM}^2}{4}\tau_P \int_{-\infty}^{\infty}  d\sigma 
 \left[ 2- e^{a\sigma} - e^{-a\sigma} \right]\cdot \nonumber \\ &\cdot
\int_0^\infty\frac{dk^+dk^-}{4\pi^2} e^{\tfrac{i}{a} [k^+ e^{-a\sigma}-k^-  e^{a\sigma}] }
 e^{-\tfrac{i}{a}[ k^+ -k^- ] }\theta(k^+k^--m^2)\theta(\Lambda^2-k^+k^-)
 \end{align}
 where we have noticed that we could do a boost $k^\pm\to k^{\pm}e^{\pm a\tau'}$ which rewrites the integrand
 as a function of $\sigma=\tau-\tau'$.  Then, we changed to coordinates $((\tau+\tau')/2,\sigma\equiv\tau-\tau')$ and, since the integrand
 did not depend of $(\tau+\tau')/2$, we integrated it to produce the overall factor $\tau_P$.  Then we take the 
 limit $\tau_P\to\infty$ in the remaining $\sigma$-integral.   
 
  We can now go to hyperbolic polar coordinates
 $k^+=\kappa e^\theta$ and $k^-=\kappa e^{-\theta}$ were $dk^+dk^-=2\kappa d\kappa d\theta$.
 We further define $z=e^\theta$, $z'=e^{a\sigma}$
  so that the integral is
 \begin{align}
 &=\frac{g_{\rm YM}^2}{8\pi^2 }(a\tau_P) \int_{m/a}^{\Lambda/a}\kappa d\kappa \int_{0}^{\infty}  \frac{dz}{z}\frac{dz'}{z'}e^{i\kappa(z-1/z) }e^{-i\kappa(z'1/z') }\left[ 2- \frac{z}{z'} -\frac{z'}{z}\right] \\
&=\frac{g_{\rm YM}^2}{8\pi^2 }(a\tau_P) \int_{m/a}^{\Lambda/a}\kappa d\kappa  8\left[K_0(2\kappa)^2+ K_1(2\kappa)^2\right] \\
&=\frac{g_{\rm YM}^2}{8\pi^2 }(a\tau_P) \left[-4\kappa K_0(2\kappa)K_1(2\kappa)\right]_{m/a}^{\Lambda/a} 
 =\frac{g_{\rm YM}^2}{4\pi^2 }(a\tau_P)\left[\ln\frac{a}{m}-\ln \frac{a}{\Lambda}\right]
 \end{align}
 where, in the last line,  we have used the small argument limits of the modified Bessel functions, kept the logarithmically divergent term and dropped a finite constant. If, instead, we had done the integral with $\Lambda\to\infty$ we would have
 obtained $\frac{g_{\rm YM}^2}{4\pi^2 }(a\tau_P) \ln\frac{a}{m}$.

  \section{Appendix: Integral I}
  \label{AppendixI}
As in appendix \ref{Appendix0}, in order to simplify the momentum integrations, we note that  integrands in $I,II,III,IV$ depend only on $k^0$ and $k^1$ and we can make the replacement
$$
\int \frac{d^3k}{(2\pi)^22|\vec k|}f(k^0,k^1)~\to~\int_0^\infty\frac{dk^+dk^-}{(2\pi)^2}f(k^0,k^1)
$$
Beyond this, we shall need an infrared regulator.  We will choose one which preserves Lorentz invariance in the $k^+$-$k^-$ plane by restricting all of the  integrals to the region $k^+k^->m^2$. Each integral will be logarithmically divergent as $m\to0$.  We shall see shortly that these logarithmic divergences cancel when the above four terms are added together. 
Beyond this, the integrals $I,II$ and $III$ have an upper cutoff, so that their integrals are also restricted to the region $k^+k^-<\Lambda^2$ where $\Lambda$ is the detector resolution. 

Integral $I$ is 
$$
\int_m^\Lambda\frac{ dk^+dk^-   }{ {(2\pi)^2} }  \frac{  \dot{\tilde x}^\mu(\tau_P/2) \dot{\tilde x}_{\mu}(-\tau_P/2) +| \dot{\tilde x}(\tau_P/2) || \dot{\tilde x}(-\tau_P/2)| } { [\dot{\tilde x}^\mu(-\tau_P/2) k_\mu ][\dot{\tilde x}^\mu(\tau_P/2) k_\mu ] } 
$$
 $$
=\int_m^\Lambda\frac{ dk^+dk^-   }{ {(2\pi)^2} }   \frac{  1-\cosh\tau_P } {[k^+e^{-\tau_P/2}+k^-e^{\tau_P/2}] [k^+e^{\tau_P/2}+k^-e^{-\tau_P/2}]} 
$$
  $$
=\int_m^\Lambda\frac{d\kappa}{\kappa}\int_{-\infty}^\infty \frac{ 2 d\theta   }{ {(2\pi)^2}}   \frac{  1-\cosh\tau_P } 
{e^{2\theta}+e^{\tau_P}+e^{-\tau_P}+e^{-2\theta} } 
$$
$$
=\frac{1}{2\pi^2}\ln\frac{\Lambda}{m} \int_0^\infty \frac {dz}{2z}   \frac{  1-\cosh\tau_P } 
{z+e^{\tau_P}+e^{-\tau_P}+1/z } 
$$
$$
=\frac{1}{4\pi^2}\ln\frac{\Lambda}{m} \int_0^\infty dz   \frac{  1-\cosh\tau_P } 
{(z+e^{\tau_P})(z+e^{-\tau_P}) } =-\frac{1}{4\pi^2}(a\tau_P)\ln\frac{\Lambda}{m}    
$$

\section{Appendix: Integral II = Integral III}
\label{AppendixII,III}
Integrals $II$ and $III$ can be shown to be identical by a change of variables.  They are
 $$
II=III= i   \int_m^\Lambda\frac{ dk^+dk^-   }{  (2\pi)^2 }    
  \int_{-\tau_P/2}^{\tau_P/2}  d\tau e^{ ik_\mu \tilde x^\mu(\tau ) }   \frac{  \left[ \dot{\tilde x}^\mu(\tau ) \dot{\tilde x}_{\mu}(-\tau_P/2) +| \dot{\tilde x}(\tau ) || \dot{\tilde x}(-\tau_P/2) |\right] } { \dot{\tilde x}^\mu(-\tau_P/2) k_\mu }   
  $$
  $$
  =-i   \int_m^\Lambda\frac{ dk^+dk^-   }{  (2\pi)^2 }    
  \int_{-\tau_P/2}^{\tau_P/2}  d\tau e^{ -ik^+e^{-\tau} -ik^-e^{\tau}} \frac{ 
  1-\cosh(\tau+\tau_P/2) } { k^+e^{\tau_P/2} +k^-e^{-\tau_P/2}   }   
 $$
 $$
 =-\frac{i}{2\pi^2} \int_m^\Lambda d\kappa\int_{-\infty}^\infty d\theta  
  \int_{-\tau_P/2}^{\tau_P/2}  d\tau e^{- i\kappa e^{\theta-\tau} -i\kappa e^{\tau-\theta}} \frac{ 
  1-\cosh(\tau+\tau_P/2) } { e^{\theta+\tau_P/2} +e^{-\theta-\tau_P/2}  } 
$$
$$
 =-\frac{i}{4\pi^2} \int_m^\Lambda d\kappa\int_{-\infty}^\infty d\theta  
  e^{- i\kappa e^{\theta} -i\kappa e^{-\theta}}  \int_{0}^{\tau_P}  d\tau\frac{ 
  2-e^{\tau}-e^{-\tau} } { e^{\theta+\tau} +e^{-\theta-\tau}  } 
$$
$$
 =\frac{i}{4\pi^2}\tau_P \int_m^\Lambda d\kappa\int_{-\infty}^\infty d\theta  
  e^{ -i\kappa e^{\theta} -i\kappa e^{-\theta}} e^{-\theta}+\ldots
  =\frac{i}{2\pi^2}\tau_P \int_m^\Lambda d\kappa K_1(2i\kappa)+\ldots 
$$
$$
 =\frac{1}{4\pi^2}(a\tau_P)\ln\frac{\Lambda}{m}+\ldots 
$$
 where ellipses are terms which grow slower than $\tau_P$ with large $\tau_P$, which we shall drop, and we have assumed that both $\Lambda/a$ and $m/a$ are so small that we can use the small argument asymptotic of the Bessel function.

\section{Appendix: Integral IV}
\label{AppendixIV}
Integral $IV$ is proportional to the integral which is
done in appendix \ref{Appendix0}.  The result is 
 \begin{align*}
 & -\frac{1}{2}\int_m\frac{ dk^+dk^-   }{ {(2\pi)^2} } 
 \int_{-\tau_P/2}^{\tau_P/2}  d\tau \int_{-\tau_P/2}^{\tau_P/2}   d\tau' 
 e^ {ik(  x(\tau )-  x(\tau')) }    \left[  \dot {\tilde x}^\mu(\tau) \dot {\tilde x}_\mu(\tau') +|\dot{\tilde x}(\tau)||\dot{\tilde x}(\tau')|\right] \\
 & = -\frac{1}{4 \pi^2}(a\tau_P)
  \ln\frac{a}{m} 
\end{align*}
where we remind the reader that we have assumed that $m<<a$. 

\end{appendix}

\section*{Acknowledgement}
This work was supported in part by NSERC. This work was performed in part at the Aspen Center for Physics, which is supported by National Science Foundation grant PHY-1607611.

\end{document}